\documentclass[preprint,12pt]{elsarticle}

\usepackage{graphicx}

\usepackage{amssymb}
\usepackage[all]{xy}
\usepackage{enumerate}
\usepackage{array}
\usepackage{amssymb}
\usepackage{amsthm}
\usepackage{amsmath}
\usepackage{amscd}
\usepackage{color}
\usepackage{hyperref}
\hypersetup{
    colorlinks,
    citecolor=green,
    filecolor=black,
    linkcolor=blue,
    urlcolor=black
}
\hypersetup{linktocpage}
\usepackage{lineno}
\usepackage{endfloat}
\usepackage{booktabs}

\theoremstyle{definition}

\theoremstyle{remark}

\def \fd4{{\textbf{FD4} \,}}

\newcommand{\K}{K_q(\tau)}

\newcommand{\lto}{\longrightarrow}

\journal{Physica A: Statistical Mechanics and its Applications}

\begin{document}

\begin{frontmatter}

\title{A comparison among some Hurst exponent approaches to predict nascent bubbles in $500$ company stocks}

\author[rvt0]{M. Fern\'andez-Mart\'{\i}nez\fnref{fn1}}
\ead{fmm124@gmail.com}
\author[rvt]{M.A S\'anchez-Granero\fnref{fn2}\corref{cor1}}
\ead{misanche@ual.es}
\author[rvt1]{Mar\'{\i}a Jos\'e Mu\~noz Torrecillas}
\ead{mjmtorre@ual.es}
\author[rvt2]{Bill McKelvey}
\ead{mckelveybill1@gmail.com}

\cortext[cor1]{Corresponding author}
\fntext[fn1]{The first author especially acknowledges the valuable support
provided by Centro Universitario de la Defensa en la Academia General del
Aire de San Javier (Murcia, Spain).}
\fntext[fn2]{The second author acknowledges the support of the Ministry of Economy and Competitiveness
of Spain, Grant MTM2012-37894-C02-01.}

\address[rvt0]{University Centre of Defence at the Spanish Air Force Academy, MDE-UPCT\\ 30720 Santiago de la Ribera, Murcia (Spain)}

\address[rvt]{Department of Mathematics, Universidad de Almer\'{\i}a, 04120 Almer\'{\i}a (Spain)}

\address[rvt1]{Department of Economics, Universidad de Almer\'{\i}a, 04120 Almer\'{\i}a (Spain)}

\address[rvt2]{Kedge Business School, Marseille, France}

\begin{abstract}
In this paper, three approaches to calculate the self-similarity exponent of a time series are compared in order to determine which one performs best to identify the transition from random efficient market behavior (EM) to herding behavior (HB) and hence, to find out the beginning of a market bubble. In particular, classical Detrended Fluctuation Analysis (DFA), Generalized Hurst Exponent (GHE) and GM2 (one of Geometric Method-based algorithms) were applied for self-similarity exponent calculation purposes. Traditionally, researchers have been focused on identifying the beginning of a crash. Instead of this, we are pretty interested in identifying the beginning of the transition process from EM to a market bubble onset, what we consider could be more interesting. The relevance of self-similarity index in such a context lies on the fact that it becomes a suitable indicator which allows to identify the raising of HB in financial markets. Overall, we could state that the greater the self-similarity exponent in financial series, the more likely the transition process to HB could start. This fact is illustrated through actual S\&P500 stocks.  
\end{abstract}


\end{frontmatter}
\section{Introduction}

\subsection*{Characterizing herding behavior}

By \cite{BIK01} Bikhchandani and Sharma cite $51$ works discussing HB pertaining to stock markets, traders, analysts, investors, mutual fund managers, and so on. They mention information-based HB and cascades as well as reputation- and computation-based HB. In \cite{HIR03} Hirshleifer and Teoh publish an even broader review, citing some $165$ works focusing on finance-relevant HB. Their `` \ldots\ focus is on convergence or divergence brought about by actual interactions between individuals'' (\cite[p. 27, their italics]{HIR03}). They refer to convergence in buy/sell behaviors and offer various reasons why rational individuals abandon their own private information about fundamental values and, instead, jump onto an HB bandwagon (\cite[p. 32]{HIR03}):
\begin{itemize}
\item \emph{Idiosyncrasy:} Loss of idiosyncrasy occurs when increasing numbers of traders follow a few initial buy/sell signals;
\item \emph{Fragility:} As traders increasingly follow fewer buy/sell signals they increasingly ignore other sources of information and lose their sensitivity to small shocks; because the herd loses its sensitivity to other sources of information a network-based cascade of buying loses its fragility and persists until a more dramatic sell-off occurs;
\item \emph{Simultaneity:} Loss of idiosyncrasy and fragility resulting when increasing numbers of traders follow the same buy/sell signals inevitably results in ``chain reactions'', ``stampedes'', or ``avalanches'' as traders increasingly jump on board the bandwagon;
\item \emph{Paradoxicality:} Traders in cascades simply ignore countervailing information as long as the herd appears to keep growing and a market keeps rising (or falling);
\item \emph{Path dependence:} Positive-feedback dynamics based on random, insignificant initial buy/sell events give rise to increasing payoffs, which then become so attractive that traders get locked-in, i.e., they are more and more reluctant to jump off the bandwagon \cite{ART89}. The positive-feedback rationale is emphasized in more recent papers as well (Hott \cite{HOT09}; Jeon and Moffett \cite{JEO10}). 
\end{itemize}
Overall, the two review articles primarily focus on identifying HB after it exists, and then argue about whether HB is rational or irrational. Research about the presence and rationales for herding in financial markets has continued since the Hirshleifer and Teoh review. More recent research is emerging that empirically connects herding to changes in stock markets (e.g., Gleason, Lee, and Mathur \cite{GLE03}; Chiang and Zheng \cite{CHI10}; Demirer, Kutan, and Chen \cite{DEM10}; Demirer and Ulussever \cite{DEM11}; Blasco, Corredor and Ferreruela \cite{BLA12}; Balcilar, Demirer and Hammoudeh \cite{BAL13}; Chen \cite{CHE13}; Kukacka and Barunik \cite{KUK13}). 
Nowhere in the finance or econophysics literature, however, do we see any empirical method proposed or used that actually identifies the transition point between EMB and HB, i.e., the beginning of HB − a recent exception is Domino \cite{DOM11}. And, as far as we know, no method is currently (publically) available that offers sound advice to stock-traders that a longer term stock-price rise (or fall) has begun and will continue for some time into the future, and thus, a good time for investing has arrived.
The structure of our paper is as follows. Section \ref{section:back} provides a mathematical sketch regarding the three approaches that we use in this study to calculate the self-similarity exponent (of financial-time series) for comparative purposes: DFA, GHE and GM2 algorithms, respectively. In Section \ref{section:emp} we conduct comparative empirical analyses of the three HB-indication methods based on financial time-series data from S\&P500 stock Index dating from 1980 to 2015. Conclusions follow.

\section{Mathematical background}\label{section:back}
The main purpose of this section is to provide a brief mathematical support regarding some concepts, notations, theoretical results and algorithms which will be necessary along this paper. According to that, the structure of this section is as follows. In Subsection \ref{sec:ssp}, we recall the key concept of self-similar process (in the classical sense of Lamperti) as well as some properties related to its increments. In this way, the class of fractional Brownian motions will be also uniquely characterized. On the other hand, the remaining Subsections \ref{sub:ghe}-\ref{sub:gm2} will contain a technical description regarding some procedures that will be applied throughout this work to calculate the self-similarity index for a given process (in particular, for actual financial series, see forthcoming Section \ref{section:emp}). They include Generalized Hurst Exponent, Detrended Fluctuation Analysis and a pretty accurate Geometric Method-based procedure.

\subsection{Self-similar processes and their increments. Fractional Brownian motions} \label{sec:ssp}
Let $(X,\mathcal{A},P)$ be a probability space and let $t\in [0,\infty)$ denote time. Thus, $\textbf{X}=\{X(t,\omega):t\geq 0\}$ is a random process (also known as a random function),
if $X(t,\omega)$ is a random variable for all $t\geq 0$ and all $\omega\in \Omega$, where $\Omega$ is a sample space. Hence, $\textbf{X}$ provides a sample function which maps $t\mapsto X(t,\omega)$ for all $\omega\in \Omega$. Consequently, $\Omega$-points do parametrize the functions $\textbf{X}:[0,\infty)\times \Omega\lto \mathbb{R}$, whereas $P$ is a probability on such a function class.\newline
In this context, let $X(t,\omega)$ and $Y(t,\omega)$ be a pair of random functions. Let us denote $X(t,\omega)\sim Y(t,\omega)$ the fact that they present the same finite joint distribution function.
Accordingly, it is said that a random process $\mathbf{X}$ is self-similar (in the Lamperti sense, see \cite{LAM62}) provided that there exists $H>0$ such that the following PL is satisfied:
\begin{equation}\label{eq:self}
X(a\, t,\omega)\sim a^{H}\, X(t,\omega),
\end{equation}
for all $a>0$ and all $t\geq 0$. Equivalently, $\mathbf{X}$ is also called a $H$-self-similar random process, as well as its corresponding parameter $H$ is known as its self-similarity index (also exponent), which is particularly called Hurst exponent in the Brownian motion case. Thus, if $X(t,\omega)$ denotes space, then  Eq. (\ref{eq:self}) states that any change of time scale $a>0$ corresponds with a change of space scale $a^{H}$. Moreover, the bigger $H$, the more dramatic becomes the change on the space variable. Further, Eq. (\ref{eq:self}) also implies a scale-invariance of the finite dimensional distribution for the corresponding random process $\textbf{X}$. 
Moreover, the increments of a random function $X(t,\omega)$ are said to be 
stationary, provided that for all $a>0$ and all $t\geq 0$, it is satisfied that
$$X(a+t,\omega)-X(a,\omega)\sim X(t,\omega)-X(0,\omega).$$
The wide class of self-similar processes having stationary increments includes the classical Brownian motions and their generalizations to stable distributions as particular cases. 
For a mathematical background regarding all of them, let us refer the reader e.g., to \cite[Section 16.2]{FAL90}.
Finally, let us recall also that a fractional Brownian motion could be characterized as the unique $H$-self-similar Gaussian process having stationary increments with parameter $H\in (0,1)$ (see \cite{SAM94}). 

\subsection{Generalized Hurst exponent}\label{sub:ghe}

The original approach contributed by English hydrologist H.E. Hurst in \cite{HUR51} was afterwards generalized in \cite{GHE91,MAN97} to test for long-range correlations on time series. The statistical performance of such approach is based on statistical $q$th-order moments for the distribution of the increments, which being modeled through a stochastic process $X(t)$, allows us to describe the evolution of (financial) time series. Accordingly, scaling properties on such series could be explored by means of the statistic that follows:
\begin{equation}\label{eq:kq2}
K_q(\tau)=\frac{\langle |X(t+\tau)-X(t)|^q \rangle}{\langle |X(t)|^q \rangle},
\end{equation}
where $\langle \cdot \rangle$ denotes the sample average over the corresponding time window for a $T$-length time series $X(t)$ and $\tau\in \{1,\ldots, \tau_{\max}\}$ is the number of considered periods (see \cite{BAR10,levy,DIM07}). Further, it is also satisfied that statistical properties for $X(t)$ scale with the time window resolution. 
Thus, the scaling behavior of $\K$ leads to the so-called \emph{generalized Hurst exponent} (GHE, herein). Observe that all the information regarding scaling properties for $X(t)$ is contained on its scaling index $H(q)$. This makes a GHE based analysis quite simple to test for scaling properties on time series.\newline
As well as it happens with other approaches to test for scaling properties on series (see e.g., MF-DFA \cite{PEN94} as well as FD algorithms \cite{FD4}), GHE provides an appropriate procedure to deal with multi-fractal (also multi-scaling) processes, namely, those ones for which the scaling exponent $H(q)$ depends on each $q$th-order moment. In such a case, different exponents characterize the scaling properties for distinct $q$th-order moments on the underlying distribution \cite{FED88,WES85}. 
Accordingly, it becomes quite natural to apply GHE to properly study unifractal processes. Thus, scaling properties of such processes become uniquely determined through a single constant $H$ which matches the self-affine exponent for the corresponding time series. Note that in such a case, a plot of $q\, H(q)$ vs. $q$ provides a straight line (equivalently, $H(q)=H$ for each $q$th-order moment, namely, the self-similarity index remains constant).\newline
In particular, if $q=1$, then GHE provides a valid tool to test for scaling properties of the absolute deviations of the process under consideration. Thus, the quantity $H(1)$ is expected to be close to the original Hurst exponent. Consequently, in this paper we will apply GHE for $q=1$ within self-similarity exponent calculation purposes.
\subsection{(Multifractal) Detrended fluctuation analysis}\label{sub:dfa}

Detrended fluctuation analysis (DFA, onwards) is a classical algorithm widely applied in financial literature to estimate the self-similarity exponent of a random process. This procedure, which was first introduced by Peng et al. in \cite{PEN94}, allows to check for correlation and scaling properties on time series (see \cite{AUS00,LIU97,DIM05}).\newline
MF-DFA, which was contributed by Kantelhardt et al. \cite{KAN02}, provides a generalization for classical DFA. A remarkable feature for this algorithm is that it could be applied to study the scaling behavior of non-stationary multifractal time series. The main idea underlying this procedure consists of taking into account deviations of $q$th-order moments from polynomial fits. In fact, notice that for $q=2$, classical DFA remains as a particular case of MF-DFA. A properly description for such an approach could be done in the following terms. Indeed, for a given time series of length equal to $T$, and for each $m=2^k<T$, let us divide it into $d=\frac{T}{m}$ non-overlapping blocks of length $m$, and for each block, let us denote $X_{m,l}$ the corresponding $l$-order polynomial fit. For our purposes, we will use linear detrending, so let us consider $l=1$, herein. Accordingly, the notation $X_{m,l}$ could be replaced merely by $X_m$. Then MF-DFA could be stated as follows:
\begin{enumerate}[(1)]
\item determine the local trend for each block $X_m$.
\item Let $\{Y(t,j): j=1,\ldots,m\}$ be the detrended process on each block, namely, the difference between the original value of the series on the block and the local trend given by the corresponding polynomial fit: $Y(t,j)=X(t,j)-X_m(t,j)$.
\item For each block, let us calculate $B_{i,q}=(\frac{1}{m}\sum_{j=1}^m Y(t,j)^2)^{\frac{q}{2}}$ for $i=1,\ldots,d$.
\item Consider the statistic $F_{m,q}=(\frac{1}{d}\sum_{i=1}^d B_{i,q})^\frac{1}{q}$, which scales following the next PL: $F_{m,q}\propto m^{H(q)}$, where $H(q)$ is the generalized Hurst exponent.
\end{enumerate}
Hence, the estimation of the generalized Hurst exponent $H(q)$ can be carried out as the slope of the linear regression which compares $\log F_{m,q}$ vs. $\log m$. 
Note that the estimator $H_{DFA}$ provides information about memory though not about distribution increments for a given random process. Further, observe that such an estimator is based on the variance (resp. the standard deviation) of either the process or its increments, but even if the variance (resp. the standard deviation) becomes infinite, this estimator still works properly (see \cite{MON99}). In particular, this means that if we apply these approach to either Brownian motions or L\'evy stable motions within different self-similarity exponents and without memory, then we will get $H_{DFA}=0.5$ for all of them.

\subsection{A geometric method-based procedure: GM2 algorithm}\label{sub:gm2}

Geometric method-based procedures (both GM1 \& GM2) were introduced in \cite{MSG08} and revisited afterwards in \cite{MFM12} to calculate the self-similarity exponent of time series. They provide two novel geometric approaches to test for scaling and correlation properties on any (financial) series.
In this paper, though, we will be focused only on GM2 approach, which will be described next.
In fact, let us assume that both the maximum and the minimum values (prices) for each period are already known. Thus, for a given time series (of $\log$ prices) whose length is equal to $T$, and for each $m=2^k<T$, let us divide it into $d=\frac{T}{m}$ non-overlapping blocks of length $m$. Hence, 
\begin{enumerate}[(1)]
\item calculate the range of each block $\mathcal{B}_i=\{B_1,\ldots,B_m\}$, namely, $R_i=\max\{B_j:j=1,\ldots,m\}-\min\{B_j:j=1,\ldots,m\}$ for each $i=1,\ldots, d$.
\item Calculate the mean of the block ranges, namely, $M_m=\frac{1}{d} \sum_{i=1}^d R_i$.
\end{enumerate}
Accordingly, the value of $H_{GM2}$ could be obtained by means of the slope of a linear regression which compares $\log M_m$ vs. $\log m$.
We would like also to point out that while DFA uses the ($\log$) return of the series, GM2 approach actually considers the corresponding $\log$ price values.


\section{Empirical tests \& Conclusions}\label{section:emp}

The main goal in this section is to apply the three algorithms described previously in Section \ref{section:back} to calculate the self-similarity index for a sample from S\&P500 stock prices, assuming that HB can be \emph{measured} through the self-similarity index. Indeed, we check that the higher the self-similarity index, the higher a ``bubble'' is or will be in progress. According to that, we test if self-similarity exponent could be used as an indicator for future performance. 

To do this, data from 1980 to present time (2015) have been considered for this application. Further, notice that $\log$ price changes have been explored in upcoming analysis. 
In this way, our analysis has been focused on the calculation of the self-similarity exponent as an indicator regarding the existence of HB and likely bubble beginnings. Recall that self-similarity index values greater than $0.5$ do indicate long-range PL correlations, due to Peng et al., \cite{PEN95}. Econophysicists call this ``persistent behavior'', so our purpose is to identify these high Hurst exponent values as an evidence of positive feedback for HB in stock markets.

Next, we explain in detail the way we have carried out our empirical study. Indeed, we have calculated the value for the self-similarity exponent of any S\&P500 stock in a given day taking into account the previous price series to that day. For instance, let us calculate the self-similarity exponent for any S\&P500 stock price series until a given day and let us also fix the window size to be equal to $128$ data.
Then let us consider the stock prices corresponding to the previous $128$ days to that day. Then the time series is changed by rolling $20$ days each time, that is, we roll the starting date $20$ days and then we consider again other $128$ data points to calculate another Hurst exponent value. Hence, we get $108$ coincident data points provided that we compare with the previous series, so note that the intervals are overlapping (we would like to point out that we have also analyzed this fact without overlapping time intervals with the obtained results being quite similar). This procedure has been repeated for any stock of the $500$ ones included in S\&P500 index. This way allows us to obtain a considerable number of observations for the self-similarity index (in fact, more than $130000$ through a $128$ window size). Once we have obtained the self-similarity exponent values for all the stocks in all the time intervals, the next step is to test the utility of such values as indicators of prices changes. Therefore, after calculating the value for the Hurst exponent in a given day and for a certain prior interval, we have calculated the $\log$ price change in the following days through the same window size for comparative purposes. As a benchmark, we have applied the $\log$ price change regardless of the value of the self-similarity $H$. Next, we gather the results by self-similarity exponent values, obtaining that the higher the self-similarity index values, the higher the mean (positive) price changes.

According to that, this methodology has been carried out for DFA, GHE and GM2 self-similarity index approaches and for different window sizes ($32, 64, 128, 256$ and $512$ days) in order to compare them and determine which one becomes more accurate to identify this kind of HB in S\&P500 stocks. Next, let us summarize the main conclusions obtained throughout this procedure. First of all, notice that the results for DFA were not meaningful due to the short length of the series. Thus, the self-similarity index values calculated by DFA are very noisy. On the other hand, both GHE and GM2 approaches do yield clear results leading to the higher the self-similarity exponent, the higher the performance of the corresponding stock.

In Tables \ref{table:1}-\ref{table:3}, it is shown (for different window sizes) the annualized expected return as a function depending on the self-similarity exponent $H$. Overall, note that for both GHE and GM2 estimators, there is a strong correlation between $H$ and the future performance of the stock in all the time frames analyzed. In addition to that, we could state that for low and very low values of $H$, the performance is worse than the benchmark, whereas for high and very high values of $H$, the performance is better than such benchmark. Moreover, both Tables \ref{table:4} and \ref{table:5} display the results for values of the self-similarity index $H$ greater than percentile $95$ and between percentiles $90$ and $95$ (that is, for extremely high values of $H$), for both GHE and GM2 algorithms, respectively. The annualized returns in such cases for all time frames and both GHE and GM2 algorithms are really noteworthy: roughly around $18-20\%$ for the highest case vs. around $13\%$ for the benchmark.




\begin{table}[htbp]
\centering
  \begin{tabular}{rrrrrr}
  \addlinespace
  \toprule
  \toprule
  \multicolumn{6}{c}{{\bf Annualized return for GHE algorithm}} \\
  \midrule
  \midrule
  {$H$ \textbf{range}} & {\bf 32} & {\bf 64} & {\bf 128} & {\bf 256} & {\bf 512} \\
  \midrule
  \bf{very low}   & $9.87\%$  & $10.38\%$ & $9.73\%$  & $10.76\%$  & $12.27\%$  \\
  \bf{low}    & $12.00\%$  & $12.83\%$ & $12.40\%$  & $11.94\%$  & $12.60\%$ \\
  \bf{normal}   & $13.74\%$  & $13.26\%$ & $12.45\%$  & $12.25\%$  & $12.48\%$ \\
  \bf{high}   & $14.16\%$  & $14.65\%$ & $14.78\%$  & $13.23\%$  & $12.61\%$ \\
  \bf{very high}   & $17.80\%$  & $15.65\%$ & $16.98\%$  & $16.06\%$  & $13.03\%$ \\
  \bf{any}   & $13.48\%$  & $13.34\%$ & $13.24\%$  & $12.84\%$  & $12.60\%$ \\
  \bottomrule
  \bottomrule
  \end{tabular}
  \caption{Annualized expected return in the next time window depending on the value of the self-similarity index $H$ (calculated in this case through GHE algorithm) in the previous time window. Note that \emph{very low} means less than percentile $20$, \emph{low} means between percentiles $20$ and $40$, \emph{normal} means between percentiles $40$ and $60$, \emph{high} means between percentiles $60$ and $80$, and finally, \emph{very high} means greater than percentile $80$.}
\label{table:1}
\end{table}

\begin{table}[htbp]
\centering
  \begin{tabular}{rrrrrr}
  \addlinespace
  \toprule
  \toprule
  \multicolumn{6}{c}{{\bf Annualized return for GM2 algorithm}} \\
  \midrule
  \midrule
  {$H$ \textbf{range}} & {\bf 32} & {\bf 64} & {\bf 128} & {\bf 256} & {\bf 512} \\
  \midrule
  \bf{very low}   & $10.43\%$  & $10.04\%$ & $10.28\%$  & $10.56\%$  & $11.67\%$  \\
  \bf{low}    & $11.66\%$  & $12.85\%$ & $10.88\%$  & $11.14\%$  & $11.87\%$ \\
  \bf{normal}   & $14.19\%$  & $12.98\%$ & $12.66\%$  & $12.21\%$  & $12.30\%$ \\
  \bf{high}   & $14.46\%$  & $13.74\%$ & $14.59\%$  & $13.24\%$  & $12.51\%$ \\
  \bf{very high}   & $16.78\%$  & $17.23\%$ & $17.98\%$  & $17.14\%$  & $14.69\%$ \\
  \bf{any}   & $13.49\%$  & $13.34\%$ & $13.24\%$  & $12.84\%$  & $12.60\%$ \\
  \bottomrule
  \bottomrule
  \end{tabular}
  \caption{Annualized expected return in the next time window depending on the value of the self-similarity index $H$ (calculated in this case through GM2 approach) in the previous time window.}
\label{table:2}
\end{table}

\begin{table}[htbp]
\centering
  \begin{tabular}{rrrrrr}
  \addlinespace
  \toprule
  \toprule
  \multicolumn{6}{c}{{\bf Annualized return for DFA algorithm}} \\
  \midrule
  \midrule
  {$H$ \textbf{range}} &  {\bf 128} & {\bf 256} & {\bf 512} \\
  \midrule
  \bf{very low}   & $13.55\%$  & $13.55\%$  & $13.60\%$  \\
  \bf{low}   & $13.05\%$  & $12.23\%$  & $12.00\%$ \\
  \bf{normal}   & $12.97\%$  & $12.67\%$  & $11.92\%$ \\
  \bf{high}    & $13.03\%$  & $12.68\%$  & $12.20\%$ \\
  \bf{very high}    & $13.61\%$  & $13.04\%$  & $13.30\%$ \\
  \bf{any}    & $13.24\%$  & $12.84\%$  & $12.60\%$ \\
  \bottomrule
  \bottomrule
  \end{tabular}
  \caption{Annualized expected return in the next time window depending on the value of the self-similarity index $H$ (calculated in this case through DFA procedure) in the previous time window.}
\label{table:3}
\end{table}

\begin{table}[htbp]
\centering
  \begin{tabular}{rrrrrr}
  \addlinespace
  \toprule
  \toprule
  \multicolumn{6}{c}{{\bf Annualized return for GHE algorithm}} \\
  \midrule
  \midrule
  {$H$ \textbf{range}} & {\bf 32} & {\bf 64} & {\bf 128} & {\bf 256} & {\bf 512} \\
  \midrule
  Percentile $90-95$   & $16.66\%$  & $16.05\%$ & $15.99\%$  & $16.70\%$  & $12.89\%$  \\
  Percentil $>95$    & $17.12\%$  & $18.50$ & $19.82\%$  & $17.46\%$  & $15.13\%$ \\
  \bottomrule
  \bottomrule
  \end{tabular}
  \caption{Annualized expected return in the next time window depending on the value of the self-similarity exponent $H$ (calculated through GHE approach) in the previous time window, for values of $H$ between percentiles 90 and 95 and values greater than percentile 95, respectively.}
\label{table:4}
\end{table}

\begin{table}[htbp]
\centering
  \begin{tabular}{rrrrrr}
  \addlinespace
  \toprule
  \toprule
  \multicolumn{6}{c}{{\bf Annualized return for GM2 algorithm}} \\
  \midrule
  \midrule
  {$H$ \textbf{range}} & {\bf 32} & {\bf 64} & {\bf 128} & {\bf 256} & {\bf 512} \\
  \midrule
  Percentile $90-95$   & $15.03\%$  & $17.78\%$ & $18.31\%$  & $17.27\%$  & $14.72\%$  \\
  Percentile $>95$    & $17.51\%$  & $20.19$ & $20.51\%$  & $20.36\%$  & $16.68\%$ \\
  \bottomrule
  \bottomrule
  \end{tabular}
  \caption{Annualized expected return in the next time window depending on the value of the self-similarity exponent $H$ (calculated through GM2 algorithm) in the previous time window.}
\label{table:5}
\end{table}


\begin{thebibliography}{99}

%
\bibitem{AUS00} M. Ausloos, 
\textit{Statistical physics in foreign exchange currency and stock markets}, 
Physica A 285 (1-2) (2000) 48-65.

\bibitem{GHE91}  A.-L. Barab\'asi and T. Vicsek,
\textit{Multifractality of self-affine fractals,}
Phys. Rev. A 44 (4) (1991) 2730-2733.
%
\bibitem{BAR12} J. Barunik, T. Aste, T. Di Matteo and R. Liu,
\textit{Understanding the source of multifractality in financial markets,}
Physica A 391 (17) (2012) 4234-4251.
%
\bibitem{BAR10} J. Barunik and L. Kristoufek,
\textit{On Hurst exponent estimation under heavy-tailed distributions,}
Physica A 389 (18) (2010) 3844-3855.
%
%
%
%
%
%
%
%

\bibitem{FAL90} K. Falconer,
\textit{Fractal Geometry. Mathematical Foundations and Applications,}
John Wiley \& Sons, Chichester, 1990.

%
%
\bibitem{FED88} J. Feder,
\textit{Fractals,}
Plenum Press, New York, 1988.
%
\bibitem{levy} M. Fern\'andez-Mart\'{\i}nez, M.A. S\'anchez-Granero and J.E. Trinidad Segovia,
\emph{Measuring the self-similarity exponent in L\'evy stable processes of financial time series,}
Physica A 392 (21) (2013) 5330-5345.
%
\bibitem{FD4} M. Fern\'andez-Mart\'{\i}nez, M.A. S\'anchez-Granero, J.E. Trinidad Segovia and I.M. Rom\'an-S\'anchez, 
\emph{An accurate algorithm to calculate the Hurst exponent of self-similar processes,}
Physics Letters A 378 (32-33) (2014) 2355-2362.
%
%
%
%
%
%
%
%
%
%
\bibitem{HUR51} H. Hurst,
\textit{Long term storage capacity of reservoirs,}
Trans. Am. Soc. Civ. Eng. 6 (1951) 770-799.
%
\bibitem{KAN02} J.W. Kantelhardt, S.A. Zschiegner, E. Koscielny-Bunde, S. Havlin, A. Bunde, H.E. Stanley, 
\textit{Multifractal detrended fluctuation analysis of nonstationary time series,}
Physica A 316 (1-4) (2002) 87-114.
%
%
%
%
%
%
%
%
\bibitem{LAM62} J.W. Lamperti,
\textit{Semi-stable stochastic processes,}
Trans. Amer. Math. Soc. 104 (1962) 62-78.
%
%
%
%
%

\bibitem{LIU97} Y. Liu, P. Cizeau, M. Meyer, C. Peng, H. Stanley, 
\textit{Correlations in economic time series}, 
Physica A 245 (3-4) (1997) 437-440.

%
%
%
%
\bibitem{MAN97} B. Mandelbrot,
\textit{Fractals and scaling in finance: discontinuity, concentration, risk,}
Springer-Verlag, New York, 1997.
%
%
\bibitem{DIM07} T. Di Matteo, 
\textit{Multi-scaling in finance,}
Quant. Financ. 7 (1) (2007) 21-36.

\bibitem{DIM05} T. Di Matteo, T. Aste and M.M. Dacorogna,
\textit{Long-term memories of developed and emerging markets: Using the scaling analysis to characterize their stage of development,}
J. Bank Financ. 29 (4) (2005) 827-851.

%
%
%
%

\bibitem{MON99} A. Montanari, M.S. Taqqu, V. Teverovsky,
\textit{Estimating long-range dependence in the presence of periodicity: an empirical study},
Math. Comput. Model. 29 (10-12) (1999) 217-228.

%
%
\bibitem{PEN94} C.-K. Peng, S.V. Buldyrev, S. Havlin, M. Simons, H.E. Stanley and A.L. Goldberger, 
\textit{Mosaic organization of DNA nucleotides,}
Phys. Rev. E 49 (2) (1994) 1685-1689.
%
%
%

\bibitem{SAM94} G. Samorodnitsky and M.S. Taqqu,
\textit{Stable non-Gaussian random processes: stochastic models with infinite variance,}
Chapman \& Hall, London, 1994.

%
%
\bibitem{MSG08} M.A. S\'anchez Granero, J.E. Trinidad Segovia, J. Garc\'{\i}a P\'erez,
\emph{Some comments on Hurst exponent and the long memory processes on capital markets}, 
Physica A (387) (2008) 5543-5551.
%
\bibitem{MFM12} J.E. Trinidad Segovia, M. Fern\'andez-Mart\'{\i}nez, M.A. S\'anchez-Granero,
\emph{A note on geometric method-based procedures to calculate the Hurst exponent},
Physica A 391 (2012) 2209-2214.
%
%
\bibitem{WES85} B.J. West, 
\textit{The Lure of Modern Science: Fractal Thinking,} 
1985 (World Scientific: Singapore).
%
%
%
%
%
%
%
%
%
%

\bibitem{ART89} W. Arthur,
\textit{Competing technologies, increasing returns, and lock-in by historical events,}
 The Economic Journal 99 (394) (1989) 116-131. 

\bibitem{BLA12} N. Blasco, P. Corredor and S. Ferreruela,
\textit{Does herding affect volatility? Implications for the Spanish stock market,}
Quantitative Finance 12 (2) (2012) 311-327.

\bibitem{BLA11} N. Blasco, P. Corredor and S. Ferreruela,
\textit{Detecting intentional herding: what lies beneath intraday data in the Spanish stock market,}
Journal of the Operational Research Society 62 (6) (2011) 1056–1066.

\bibitem{DEM11} R. Demirer and T. Ulussever,
\textit{Investors herds and oil prices: Evidence from GCC stock markets,}
Working paper, Department of Economics and Finance, Southern Illinois University, Edwardsville, IL (2011).

\bibitem{DOM11} K. Domino,
\textit{The use of the Hurst exponent to predict changes in trends on the Warsaw stock exchange,}
Physica A 390 (1) (2011) 98–109.

\bibitem{GAO13} Y. Gao, S. Cai, L. L \"{u} and B. Wang,
\textit{Evolutionary model on market ecology of investors and investments,}
 Physica A (16) (2013) 3385–3391.

\bibitem{YOO08} S.-H. Yook and Y. Kim,
\textit{Herd behavior in weight-driven information spreading models for financial market,}
Physica A 387 (26) (2008) 6605-6612.

\bibitem{WEL00} I. Welch,
\textit{Herding among security analysts,}
Journal of Financial Economics 58 (3) (2000) 369-396.

\bibitem{SHI90} R.J. Shiller,
\textit{Speculative Prices and Popular Models,}
Journal of Economic Perspectives 4 (2) (1990) 55-65.

\bibitem{SHI00} R.J. Shiller,
\textit{Irrational Exuberance,}
Princeton University Press, Princeton, NJ, 2000.

\bibitem{SHI02} R.J. Shiller,
\textit{Bubbles, human judgment, and expert opinion,}
Financial Analysts Journal 58 (3) (2002) 18-26.

\bibitem{SCH90} D.S. Scharfstein and J.C. Stein,
\textit{Herd Behavior and Investment,}
The American Economic Review 80 (3) (1990) 465-479.

\bibitem{NEW05} M. Newman,
\textit{Power laws, Pareto distributions and Zipf's law,}
Contemporary Physics 46 (5) 2005 323-351.

\bibitem{BOU00} R. Cont and J.-P. Bouchaud,
\textit{Herd behavior and aggregate fluctuations in financial markets,}
Macroeconomic Dynamics 4 (2) (2000) 170-196.

\bibitem{DEM10} R. Demirer, A. Kutan and C. Chen,
\textit{Do investors herd in emerging stock markets?: Evidence from the Taiwanese market,}
Journal of Economic Behavior \& Organization 76 (2) (2010) 283-295.


\bibitem{BIK92} S. Bikhchandani, D. Hirshleifer and I. Welch,
\textit{A Theory of Fads, Fashion, Custom, and Cultural Change as Informational Cascades,}
Journal of Political Economy 100 (5) (1992) 992-1026.

\bibitem{BIK01} S. Bikhchandani and S. Sharma,
\textit{Herd behavior in financial markets,}
IMF Staff Papers 47 (3) (2001) 279-310.

\bibitem{BAR02} A.-L. Barabási,
\textit{Linked: The New Science of Networks,}
Perseus, Cambridge, MA, 2002.

\bibitem{BAN92} A.V. Banerjee,
\textit{A simple model of herd behavior,}
The Quarterly Journal of Economics 107 (3) (1992) 797-817. 

\bibitem{BAL13} M. Balcilar, R. Demirer and S. Hammoudeh,
\textit{Investor herds and regime-switching: Evidence from Gulf Arab stock markets,}
Journal of International Financial Markets, Institutions and Money 23 (2013) 295-321.

\bibitem{BRO08} J. Brown, Z. Ivkovi\'{c}, P. Smith and S. Weisbenner, 
\textit{Neighbors Matter: Causal Community Effects and Stock Market Participation,}
The Journal of Finance 63 (3) (2008) 1509-1531.

\bibitem{CAJ06} D. Cajueiro and R. De Camargo, 
\textit{Minority game with local interactions due to the presence of herding behavior,}
Physics Letters A 355 (4-5) (2006) 280-284.

\bibitem{CHA00} E.C. Chang, J.W. Cheng and A. Khorana,  
\textit{An examination of herd behavior in equity markets: An international perspective,}
Journal of Banking \& Finance 24 (10) (2000) 1651-1679.

\bibitem{CHE13} T. Chen,  
\textit{Do Investors Herd in Global Stock Markets?,}
Journal of Behavioral Finance 14 (3) (2013) 230-239.

\bibitem{CHI10} T.C. Chiang and D. Zheng,  
\textit{An empirical analysis of herd behavior in global stock markets,}
Journal of Banking \& Finance 34 (8) (2010) 1911-1921.

\bibitem{DEV96} A. Devenow and I. Welch, 
\textit{Rational herding in financial economics,}
European Economic Review 40 (3-5) (1996) 603-615.

\bibitem{BIL14} N.Escoffier and B. McKelvey, 
 Using ``Crowd-Wisdom Strategy'' to Reduce Market Failure: Proof-of-Concept from the Movie Industry, in: R. DeFillippi and P. Wikstr\"{o}m, (Eds.), International Perspectives on Business Innovation in the Creative Industries, Edward Elgar, Cheltenham, UK, 2014, pp. 200–222.

\bibitem{FAM65} Eugene F. Fama, 
\textit{The Behavior of Stock-Market Prices,}
The Journal of Business 38 (1) (1965) 34-105.

\bibitem{FAM70} Eugene F. Fama, 
\textit{The Behavior of Stock-Market Prices,}
The Journal of Finance 25 (2) (1970) 383-417.

\bibitem{FRO92} J. Stein, K. Froot and D. Scharfstein, 
\textit{Herd on the Street: Informational Inefficiencies in a Market with Short-Term Speculation,}
Journal of Finance 47 (4) (1992) 1461-1484.

\bibitem{GLE03} K. Gleason, C. Lee and I. Mathur,
\textit{Herding Behavior in European Futures Markets,}
Finance Letters 1 (1) (2003) 5-8. 

\bibitem{GOL07} B. Goldfarb, D. Kirsch and D. Miller,
\textit{Was There Too Little Entry During the Dot Com Era?,}
Journal of Financial Economics 86 (1) (2007) 100-144.

\bibitem{HIR06} D. Hirshleifer, A. Subrahmanyam and S. Titman, 
\textit{Feedback and the success of irrational investors,}
Journal of Financial Economics 81 (2) (2006) 311–338.

\bibitem{HIR03} D. Hirshleifer and S. Teoh, 
\textit{Herd Behaviour and Cascading in Capital Markets: a Review and Synthesis,}
European Financial Management 9 (1) (2003) 25-66.

\bibitem{HON04} H. Hong, J. Kubik and J. Stein, 
\textit{Social Interaction and Stock-Market Participation,}
The Journal of Finance 59 (1) (2004) 137-163.

\bibitem{JAI87} A. Jain and S. Gupta, 
\textit{Some Evidence on "Herding" Behavior of U.S. Banks,}
Journal of Money, Credit and Banking 19 (1) (1987) 78-89.

\bibitem{KUK13} J. Kukacka and J. Barunik, 
\textit{Behavioural breaks in the heterogeneous agent model: The impact of herding, overconfidence, and market sentiment,}
Physica A 392 (23) (2013) 5920-5938.

\bibitem{LO88} A. Lo and A. MacKinlay, 
\textit{Stock Market Prices Do Not Follow Random Walks: Evidence from a Simple Specification Test,}
Review of Financial Studies 1 (1) (1988) 41-66.

\bibitem{LO02} A. Lo and A. MacKinlay, 
\textit{A Non-Random Walk Down Wall Street,}
Princeton University Press, Princeton, NJ, 2002.

\bibitem{MAS07} J. Maskawa,
\textit{Stock price fluctuations and the mimetic behaviors of traders,}
Physica A 382 (1) (2007) 172-178.

\bibitem{PEN95} C. Peng, S. Havlin, E. Stanley and A. Goldberger,
\textit{Quantification of scaling exponents and crossover phenomena in nonstationary heartbeat time series,}
Chaos 5 (1) (1995) 82-87.

\bibitem{SHL00} A. Shleifer, 
\textit{Inefficient Markets: An Introduction to Behavioral Finance,}
Clarendon Lectures in Economics, Oxford University Press, 2000.

\bibitem{HOT09} C. Hott,
\textit{Herding behavior in asset markets,}
Journal of Financial Stability 5 (1) (2009) 35-56.

\bibitem{JEO10} J.Q. Jeon and C.M. Moffett, 
\textit{Herding by foreign investors and emerging market equity returns: Evidence from Korea,}
International Review of Economics \& Finance 19 (4) (2010) 698-710.

\bibitem{DYE09} J.R.G Dyer, A. Johansson, D. Helbing, I.D. Couzin and J. Krause, 
\textit{Leadership, consensus decision making and collective behaviour in humans,}
Phylosophical Transactions of the Royal Society B 364 (1518) (2009) 781-789.

\bibitem{BAD10} M. Baddeley, 
\textit{Herding, social influence and economic decision-making: socio-psychological and neuroscientific analyses,}
Phylosophical Transactions of the Royal Society B 365 (1538) (2010) 281-290.

\bibitem{TRU94} B. Trueman, 
\textit{Analyst Forecasts and Herding Behavior,}
The Review of Financial Studies 7 (1) (1994) 97-124.

\bibitem{CHR95} W.G. Christie and R.D. Huang, 
\textit{Following the Pied Piper: Do Individual Returns Herd around the Market?,}
Financial Analysts Journal 51 (4) (1995) 31-37.

\bibitem{GRI94} M. Grinblatt, S. Titman and R. Wermers, 
\textit{Momentum Investment Strategies, Portfolio Performance, and Herding: A Study of Mutual Fund Behavior,}
The American Economic Review 85 (5) (1994) 1088-1105.

\bibitem{AVE98} C. Avery and P. Zemsky, 
\textit{Multidimensional Uncertainty and Herd Behavior in Financial Markets,}
The American Economic Review 88 (4) (1998) 724-748.

\bibitem{BRU01} M.K. Brunnermeier,
\textit{Asset Pricing under Asymmetric Information: Bubbles, Crashes, Technical Analysis, and Herding,}
Oxford University Press, Oxford, UK, 2001.

\bibitem{PAR05} W.D. Parker, and R.R. Prechter, 
\textit{Herding: An Interdisciplinary Integrative Review from a Socionomic Perspective,}
Available at SSRN: \url{http://ssrn.com/abstract=2009898} or \url{http://dx.doi.org/10.2139/ssrn.2009898}

\bibitem{CAM06} D. Bernhardt, M. Campello and E. Kutsoati, 
\textit{Who herds?,}
Journal of Financial Economics 80 (3) (2006) 657-675.

\bibitem{CLA06} J. Clarke and A. Subramanian, 
\textit{Dynamic forecasting behavior by analysts: Theory and evidence,}
Journal of Financial Economics 80 (1) (2006) 81-113.

\bibitem{ROO06} L. Rook, 
\textit{An Economic Psychological Approach to Herd Behavior,}
Journal of Economic Issues 40 (1) (2006) 75-95.

\bibitem{PRO12} J.M. Prosad, S. Kapoor and J. Sengupta, 
\textit{An Examination of Herd Behavior: An Empirical Evidence from Indian Equity Market,}
International Journal of Trade, Economics and Finance 3 (2) (2012) 154-157.

\bibitem{SAU13} S.N. Bhaduri and S.D. Mahapatra, 
\textit{Applying an alternative test of herding behavior: A case study of the Indian stock market,}
Journal of Asian Economics 25 (2013) 43-52.

\bibitem{LAI13} Y.-W. Laih and Y.-S. Liau, 
\textit{Herding Behavior during the Subprime Mortgage Crisis: Evidence from Six Asia-Pacific Stock Markets,}
International Journal of Economics and Finance 5 (7) (2013) 71-84.

\bibitem{OUA13} M. Ouarda, A. el Bouri and O. Bernard,
\textit{Herding Behavior under Markets Condition: Empirical Evidence on the European Financial Markets,}
International Journal of Economics and Financial Issues 3 (1) (2013) 214-228.

\bibitem{TAN08} L. Tan, T.C. Chiang, J.R. Mason and E. Nelling,
\textit{Herding behavior in Chinese stock markets: An examination of A and B shares,}
Pacific-Basin Finance Journal 16 (1-2) (2008) 61-77.

\bibitem{BAC14} L. Bachelier,
\textit{Le jeu, la chance et le hasard,}
E. Flammarion, 1914.




%
%
%
%
%
%


\end{thebibliography}
\end{document}